\documentclass[14pt]{article}
\usepackage{graphicx}
\boldmath
\begin{document}

\begin{center}
{\bf\LARGE Superlong GRBs \\
\vspace{5mm}
\rm 
\Large Ya. Yu. Tikhomirova$^1$ and   B. E. Stern$^{2,1}$\\
\rm
\it\small
$^1$  Astro Space Centre of Lebedev Physical Institute, 
Moscow, Russia\\
$^2$ Institute for Nuclear Research, Moscow, Russia
}
\end{center}
\rm

{\bf Abstract—} We searched for anomalously long GRBs 
(GRBs) in the archival records of the 
Burst and Transient Sources Experiment (BATSE). 
Ten obvious superlong ($>$500 s) GRBs 
with almost continuous emission episodes were found. 
Nine of these events are known from the BATSE catalog, 
but five have no duration estimates; 
we found one burst for the first time. 
We also detected events with emission episodes 
separated by a long period of silence (up to 1000 s) 
with a total duration of 1000--2000 s. 
In the latter case, we cannot reach 
an unequivocal conclusion about a common origin 
of the episodes due to the BATSE poor angular resolution. 
However, for most of these pairs, 
the probability of a coincidence of independent GRBs 
is much lower than unity, and the probability that 
all of these are coincidences is $\sim 10^{-8}$. 
All of the events have a hardness ratio 
(the ratio of the count rates in different energy channels) 
typical of GRBs, and their unique duration is unlikely 
to be related to their high redshifts. 
Superlong bursts do not differ in their properties 
from typical long ($>$2 s) GRBs. We estimated the fraction 
of superlong GRBs ($>$500 s) among the long ($>$2 s) GRBs 
in the BATSE sample with fluxes up to 0.1 $ph cm^{-2} s^{-1}$ 
to be between 0.3 and 0.5\%, which is higher than 
the estimate based on the BATSE catalog. 

\section{INTRODUCTION}
   At present, it is well known that the events 
beginning as cosmic GRBs (GRBs), 
isolated nonrecurrent bursts with very different 
light-curve profiles commonly observed in the energy range 
from tens to hundreds of keV, are subsequently followed by 
optical and X-ray emission (Van Paradijs et al. 2000). 
The following important question remains: 
how long do these events last in the gamma-ray range (GRBs proper)? 
There is no universally accepted model of the physical GRB 
emission mechanism (although the question of their association 
with a certain class of supernovae seems to have been solved), 
and the existence of anomalously long GRBs can impose 
significant constraints on the model of this mechanism.

   BATSE was the most sensitive among the experiments 
in which GRBs were observed (Fishman 1989). 
Furthermore, this was an experiment in which 
a continuous all-sky monitoring was performed 
in the energy range 25 keV--1 MeV during 
a uniquely long period of 9.1 yr (1991-2000). 
Before the beginning of this experiment, 
GRBs were considered to be a unified class of events 
with a duration of up to about 100 s (Mazets et al. 1981). 
According to the BATSE catalog (see Table 1), 
more than a hundred events with a duration of 100--500 s, 
three events with a duration of 500--1000 s, 
and one event with a record duration of 1300 s 
were detected during the BATSE experiment 
among $\sim$2700 bursts.

   In addition, a subclass of short bursts with 
a duration of up to $<$2 s with harder spectra 
that account for a quarter of the total number 
of BATSE GRBs was revealed.

   There are several types of BATSE records, 
in particular, records of the continuous all-sky monitoring 
by eight BATSE detectors with a time resolution 
of 1024 ms in four energy channels (DISCLA)
and with a time resolution of 2048 ms in 16 channels 
(CONT) in the energy range 25--1000 keV.
   Recently, Connaughton (2002) searched for 
long-lived GRB emission in the CONT records. 
Having added up the episodes of records 
from the onset of bursts to thousands of seconds 
for hundreds of events, she found that the signal 
extends to $\sim$1000 s and is observable even longer. 
Since the bursts with an emission episode 
widely separated from the main peak were excluded 
from the analysis, this suggests that the emission 
of long GRBs on time scales of $\sim$1000 s 
could be their common feature.

   Whereas Connaughton (2002) applied a statistical approach, 
we used an individual approach and searched for 
anomalously long bursts in the BATSE DISCLA records.

\begin{table}
\caption{ Statistics of superlong GRBs found from BATSE data.}
\vspace{0.4cm}
\begin{center}
\begin{tabular}{|c|c|c|}
\hline
\small
&\multicolumn{2}{|c|}{ Number of bursts with T $>$ 500 s}\\
GRB sample&500-1000s&1000-2000s\\
\hline
&&\\
BATSE catalog$^*$& 3 & 1 \\
(according to table of durations)&&\\
UC catalog by Stern and Tikhomirova$^{**}$& 5 & 1 \\
(search in BATSE DISCLA records)&&\\
Superlong candidate pairs in UC catalog& 1 & 6 \\
&&\\
\hline
\multicolumn{3}{c}{$^*$\footnotesize{
Accessible at http://gammaray.msfc.nasa.gov/batse/grb/catalog/current/}}\\
\multicolumn{3}{c}{$^{**}$\footnotesize{
 Accessible at http://www.astro.su.se/groups/head/grb\_archive.html
}}\\
\end{tabular}
\end{center}
\end{table}

\section{THE SEARCH FOR SUPERLONG ($>$500 s) GRBs}
   We scanned the records of the photon count rate 
in each of the eight BATSE detectors 
in two of the four energy channels, 50--100 and 100--300 keV 
(which coincides with the recorded spectral range in BATSE itself) 
with a time resolution of 1024 ms over the entire period 
of BATSE observations, 9.1 yr. The scanning was performed 
as part of our search for nontriggered GRBs 
(Stern et al. 2001, 2002). However, special attention was paid 
to the search for anomalously long bursts, 
including those with unusual profiles; 
additional runs of records with variations of 
the main search criteria were carried out.

   By nontriggered bursts we mean events 
whose parameters did not satisfy the triggering criteria 
(most frequently due to a low intensity) 
in the main event-recording procedure; 
therefore, these events were missed and were 
not recorded during the experiment itself. 
However, being statistically significant and 
reliably classifiable, these can be found in the BATSE records. 
In contrast to the BATSE catalog, which contains GRBs 
with a peak flux down to 0.2 $ph cm^{-2} s^{-1}$ 
(Paciesas et al. 1999), the search for nontriggered events 
by Stern et al. (2001) revealed bursts with a peak flux down
to $\sim$ 0.1 $ph cm^{-2} s^{-1}$.

   Our scans of the DISCLA records revealed 
a total of 3906 GRBs (see the catalog by Stern and Tikhomirova 
and Stern et al. 2001); 
1838 of these are new, previously unknown, nontriggered events, 
and 2068 are GRBs known from the BATSE catalog. 
Among all the detected GRBs, ten are obvious superlong events 
with a duration longer than 500 s (see Fig. 1 and Table 1).
Nine of these are known triggered events from the BATSE catalog, 
but only four of these were known as superlong bursts, 
according to the table of durations in the catalog. 
Our estimates for the five remaining GRBs and 
one nontriggered GRB found for the first time are given 
in Table 2. Only one of these six bursts, GRB 910425, 
has an estimate of its duration in the BATSE catalog, 
t90 $=$ 430 s, which is clearly indicative of the loss 
of episodes and underestimation of the event duration.

Note that the parameter T90, the time in which
from 5 to 95\% of the total burst flux is radiated, 
is commonly used as the GRB duration. 
Here, however, we study the maximum duration 
of the gamma-ray emission, and it is inappropriate 
to use this parameter. Therefore, in this paper, 
we use mainly an estimate of the total burst duration 
by visually determining the beginning and the end of the event. 
In addition, it is more difficult to estimate T90 
for superlong $>$500 s bursts in the BATSE DISCLA records 
than for $<$ 100 s bursts, since the fluctuations of the high, 
strongly variable background are difficult to take into account 
on long time scales. Breaks are also occasionally encountered 
in the records, making this estimate impossible.

   The longest of the six bursts whose duration was 
not estimated or underestimated in the BATSE catalog, 
GRB971208 (see Fig. 1), is known from Connaughton et al. (1997) 
and is unique in its own way. It has a classical 
single-peak profile with a rapid rise and an exponential decay. 
However, it has both an anomalously long rise time $\sim$70 s, 
and an anomalously long tail. This burst has a smooth profile 
and the longest duration, $\sim$1000 s, among the bursts 
with a classical profile (the second longest burst with 
a similar profile has a duration of $\sim$250 s). 
After $\sim$1000 s, the record is interrupted for $\sim$500 s, 
which does not allow us to estimate the duration of this burst 
more accurately.

   An important point is that for an anomalously long burst, 
its episodes can be recorded as independent events, 
particularly if the interval of silence
between the episodes is much longer than their duration. 
Therefore, we checked all of the GRBs found 
in the BATSE records for the presence of correlated pairs,
which could be episodes of anomalously long events. 
We searched for bursts that were separated by no more than 
2000 s and that had overlapping error regions. We
found nine pairs of such events.

   In view of the specific properties of gamma-ray detectors, 
the localization accuracy for these GRBs is very low, $\sim$10$^o$. 
Therefore, we cannot reliably determine whether each of the pairs 
found is a superlong burst from the same source or independent
bursts from different sources. Invoking data from the Ulysses 
interplanetary spacecraft (Hurley 2004), which operated 
simultaneously with BATSE in the same spectral range, 
for a triangulation attempt did not allow us to clarify 
the situation due to the low sensitivity 
(in all cases, Ulysses does not see one of the episodes), 
and we can estimate the confidence level only theoretically.

\begin{table}
\small
\caption{ Obvious superlong ($>$500 s) GRBs found in 
the BATSE DISCLA archival records.}
\vspace{0.4cm}
\begin{center}
\begin{tabular}{|c|c|c|c|c|c|c|c|c|}
\hline
Date&
Seconds&
Catalog&
T,s&
$\alpha$, deg&
$\delta$, deg&
R$^a$&
F$^b_{peak}$,&
HR$^c$\\
(TJD)&of Day&
entry&
&&&
deg&
$_{ph cm^{-2} s^{-1}}$&
$_{2+3/1}$\\
\hline
\multicolumn{9}{|c|}{\bf
Superlong, with duration not (under)estimated in BATSE catalog}\\
{\bf 971208 }& {\bf 28085 }& {\bf UC 10790a }& {\bf 1000} &{\bf 355.8 }
& {\bf 76.2 }& {\bf 0.4 } & {\bf 2.0 }& {\bf 2.36 } \\
(10790) & 28092 & BATSE 6526 & None & 356.5 & 77.9 & 1.2 & 1.8 &  \\
&&&&&&&&\\
{\bf 960425 }& {\bf 1102 }& {\bf UC 10198a }& {\bf 740} &{\bf 64.3 }
& {\bf 45.5 }& {\bf 9.7 } & {\bf 0.45 }& {\bf 2.38 } \\
(10198) & 1113 & BATSE 5446 & None & 59.9 & 40.8 & 5.1 & 0.44 &  \\
&&&&&&&&\\
{\bf 981104 }& {\bf 8114 }& {\bf UC 11121a }& {\bf 730} &{\bf 108.5 }
& {\bf 11.5 }& {\bf 3.2 } & {\bf 1.27 }& {\bf 2.26} \\
(11121) & 8160 & BATSE 7188 & None & 104.3 & 12.4 & 2.8 & None   \\
&&&&&&&&\\
{\bf 990123 }& {\bf 5838 }& {\bf UC 11201a }& {\bf 660} &{\bf 87.7 }
& {\bf -4.4 }& {\bf 9.7 } & {\bf 0.33 }& {\bf 4.1 } \\
(11201) & \multicolumn{8}{|c|}{nontriggered} \\
&&&&&&&&\\
{\bf 910425 }& {\bf 20208 }& {\bf UC 8371c }& {\bf 610} &{\bf 341.8 }
& {\bf 25.0 }& {\bf 7.9 } & {\bf 0.45 }& {\bf 3.11 } \\
(8371) & 20253 & BATSE 110 & 430$^*$ & 335.9 & 25.8 & 4.8 & 0.37 &  \\
&&&&&&&&\\
{\bf 981219 }& {\bf 34133 }& {\bf UC 11166b }& {\bf 530} &{\bf 216.0 }
& {\bf -44.6 }& {\bf 5.5 } & {\bf 0.88 }& {\bf 3.8 } \\
(11166) & 33948 & BATSE 7270 & None & 218.9 & -43.6 & 2.0 & 0.9 &  \\
\hline
\multicolumn{9}{|c|}{\bf
Superlong, already known from BATSE catalog }\\
{\bf 970315 }&{\bf  80021 }& {\bf UC 10522d }&{\bf 1360}& {\bf 129.9 }
& {\bf -52.6 }& {\bf 3.5 } & {\bf 0.66} & 3.8\\
(10522) & 80022 & BATSE 6125 & 1307. & 130.6 & -52.4 & 0.99 & None & \\
&&&&&&&&\\
{\bf 950305 }&{\bf  54303 }& {\bf UC 9781b }&{\bf 930}& {\bf 195.5 }
& {\bf -12.2 }& {\bf 0.9 } & {\bf 8.17} & {\bf 3.05}\\
(9781) & 54305 & BATSE 3458 & 674. & 197.2 & -11.1 & 0.3 & 8.1 & \\
&&&&&&&&\\
{\bf 971029 }&{\bf  22424 }& {\bf UC 10750b }&{\bf 730}& {\bf 70.7 }
& {\bf -40.5 }& {\bf 1.5 } & {\bf 1.73} & {\bf 3.27}\\
(10750) & 22429 & BATSE 6454 & 616. & 66.9 & -45.0 & 1.3 & 1.62 & \\
&&&&&&&&\\
{\bf 950509 }&{\bf  83719 }& {\bf UC 9846b }&{\bf 700}& {\bf 41.6 }
& {\bf 27.0 }& {\bf 6.8 } & {\bf 1.26} & {\bf 1.71}\\
(9846) & 83766 & BATSE 3567 & 590. & 44.4 & 22.9 & 3.2 & 1.05 & \\
&&&&&&&&\\
\hline
\multicolumn{9}{l}{$^a$\footnotesize{ Radius of the error circle.}}\\
\multicolumn{9}{l}{$^b$\footnotesize{ Peak flux in the BATSE 
spectral channels 2--3 (50--300 keV).}}\\
\multicolumn{9}{l}{$^c$\footnotesize{Hardness as the ratio of 
the count rates in spectral channels 2--3 (50--300 keV) 
and 1 (25--50 keV).}}\\
\multicolumn{9}{l}{$^*$\footnotesize{ $t_{90}$}}
\end{tabular}
\end{center}
\end{table}

\begin{figure}
\includegraphics[width=10cm, height=8cm]{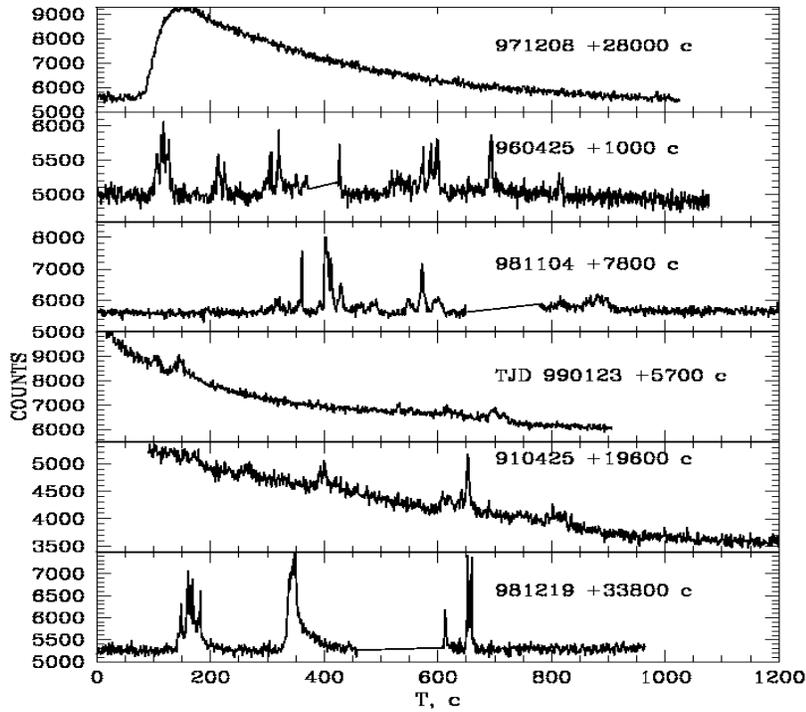}
\includegraphics[width=10cm, height=8cm]{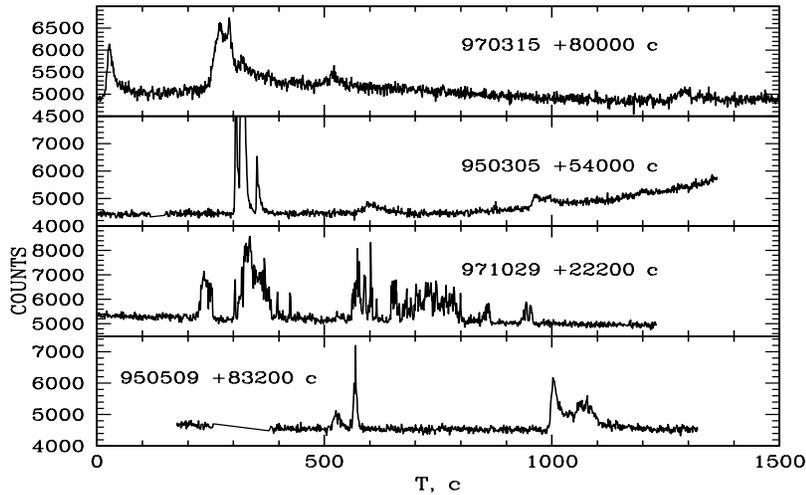}
\caption{
 Fragments of the BATSE DISCLA records containing 
obvious superlong($>$500 s) GRBs: 
(a) events whose duration was not estimated or 
underestimated in the BATSE catalog; 
and (b) superlong events known from the BATSE catalog. 
The sum of the photon count rates in two of the eight 
BATSE detectors in which the flux was at a maximum 
(the second and third energy channels, 50--300 keV) is shown. 
The date and duration are indicated for each burst.}
\end{figure}

\begin{figure}
\includegraphics[width=12cm, height=12cm]{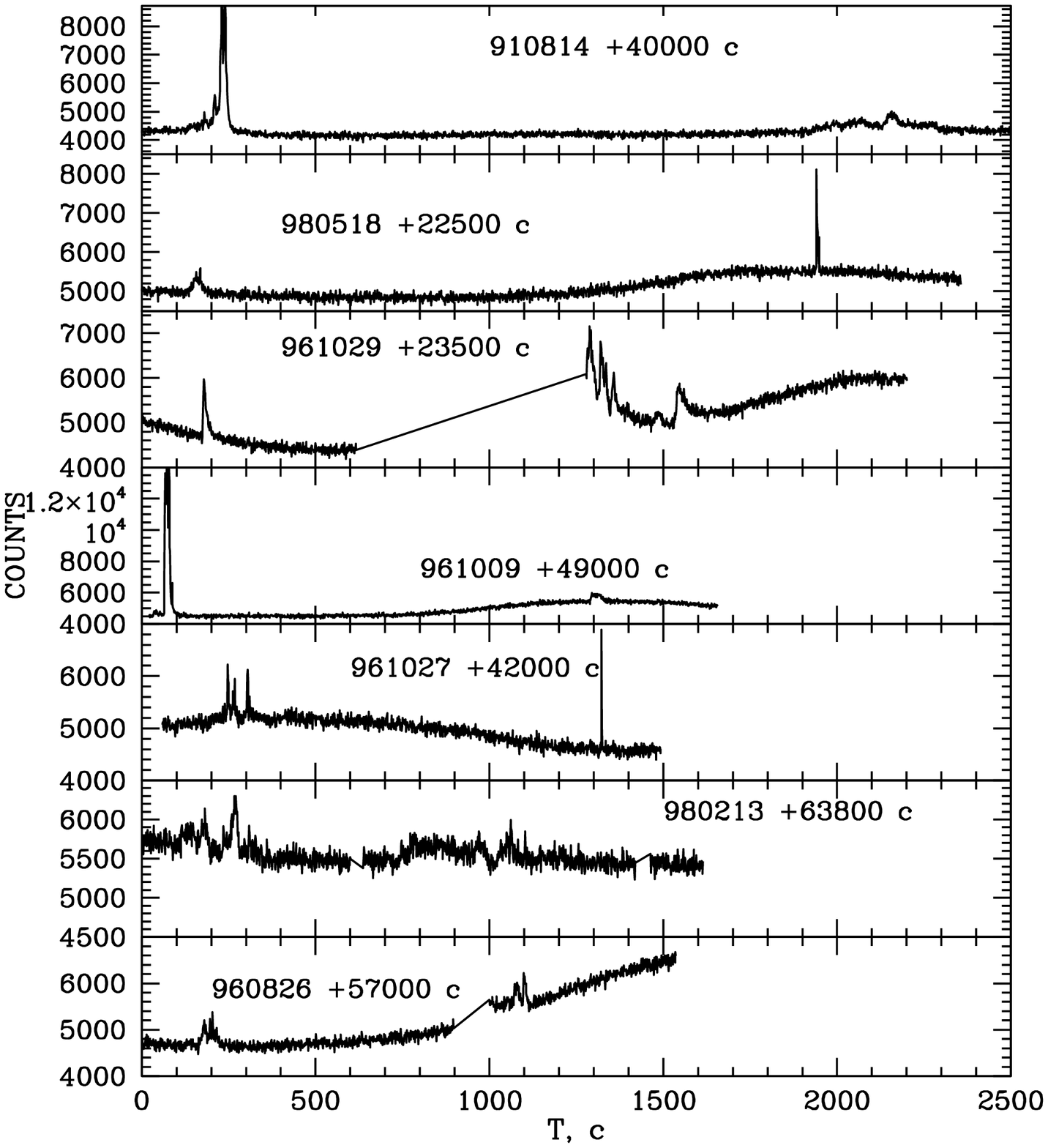}
\caption{ Fragments of the BATSE DISCLA records 
containing the detected pairs of GRBs that 
could be single superlong bursts. The sum of 
the photon count rates in two of the eight BATSE detectors 
in which the flux was at a maximum 
(the second and third energy channels, 50--300 keV) is shown. 
The date and duration are indicated for each event.
}
\end{figure}

   For each of the nine pairs, we estimated the expectation 
of the observation of two independent bursts in the sample 
of detected GRBs with a separation in angle and time no larger 
than a given value and with fluxes no lower than given values:
$$
P = {(1-cos\sigma) \over 2} \Delta T \cdot R_{GRB} N(F_p),
$$
where $\sigma$ is the angular separation between the episodes, 
$\Delta T$ is the time interval between them, 
$R_{GRB}$ is the GRB detection rate 
($R_{GRB} = N/(T_{obsr})) = 1.75 \cdot 10^{-5} s^{-1}, N = 3400$ 
is the total number of detected long GRBs, $T_{obs} = 9.1 yr$ 
is the BATSE operation time,  $\eta = 0.7$ is the exposure factor 
that includes the gaps in the data and the periods of 
a high background), and $N(F_p)$ is the number of 
detected bursts with fluxes no lower than a given value. 
We introduced the last factor in such a form, 
because the probability of detecting both episodes 
(if these have a common origin) depends on 
the brightness of the event.

   For two pairs, this number was $>$1, and these were excluded 
from the analysis. We consider the remaining seven pairs 
as candidates for superlong GRBs. Their duration 
from the beginning of the first episode to the end 
of the last episode ranges from $\sim$ 900 s to $\sim$ 2300 s. 
The probability of a coincidence of seven such pairs 
of independent bursts in the sample is roughly equal 
to the product of their expectations, $\sim10^{-8}$; 
i.e., at least some of the pairs found could be superlong bursts 
with a high probability.

   The light curves of the detected pairs are shown in Fig. 2, 
and some of their parameters, including the expectation 
of a coincidence, are listed in Table 3. 
Both events were triggered, recorded during BATSE 
and included in the catalog as independent events, 
only in two pairs. In two of the remaining pairs, 
both events were nontriggered; 
these were not recorded during BATSE 
(probably because of their low intensity). 
We detected them when scanning the BATSE DISCLA records. 
In the three remaining events, one burst is nontriggered 
and another is triggered.

   It should be noted that the BATSE observational algorithm 
was not intended for the detection of superlong GRBs: 
from the detection time immediately after the accumulation 
of limited information, data were

\begin{table}
\small
\caption{ Pairs of GRBs, candidates for single superlong
($>$500 s) bursts found in the BATSE DISCLA archival records.  }
\vspace{0.4cm}
\begin{center}
\begin{tabular}{|c|c|c|c|c|c|c|c|c|}
\hline
Date (TJD)&
Seconds&
Catalog&
$T_{total}$, s&
$\alpha$, deg&
$\delta$, deg&
R&
$F_{peak}$, &
HR\\
&of day& entry&$T_{event}$&
&&deg& $_{ph cm^{-2} s^{-1}}$&
$_{2+3/1}$\\
\hline
\multicolumn{4}{|r}{\bf 2300}&\multicolumn{5}{l|}{\bf P=0.06}\\
{\bf 910814 }& {\bf 40132 }& {\bf UC 08482c }& {\bf 72.}&{\bf 62.2 }& {\bf 46.8
}
& {\bf 0.8 }& {\bf 4.29 }& {\bf 3.46 }     \\
(08482) & 40180 & BATSE 676 & 78. &58.0 & 45.2 & 1.0 & 4.08 & \\
& {\bf 41941 }& {\bf UC 08482d }& {\bf 324.} &{\bf 60.0 }& {\bf 35.3 }& {\bf 3.1
 }
& {\bf 0.58 }& {\bf 2.12 }\\
&-&-&-&-&-&-&-&\\
\multicolumn{4}{|r}{\bf 1810}&\multicolumn{5}{l|}{\bf P=0.03}\\
{\bf 980518 }& {\bf 22649 }& {\bf UC 10951a }& {\bf 42.} &{\bf 160.3 }& {\bf -44
.5 }
& {\bf 11.1 }& {\bf 0.35 }& {\bf 1.69 }\\
(10951) &-&-&-&-&-&-&-& \\
& {\bf 24441 }& {\bf UC 10951b }& {\bf 8.} &{\bf 164.2 }& {\bf -41.9 }& {\bf 5.4
 }
& {\bf 1.34 }& {\bf 2.4 }\\
& 24441 & BATSE 6762 & 8.& 162.1 & -42.5 & 2.0 & 1.46 & \\
\multicolumn{4}{|r}{\bf 1500}&\multicolumn{5}{l|}{\bf P=0.02}\\
{\bf 961029 }& {\bf 23676 }& {\bf UC 10385b }& {\bf 24.} &{\bf 62.1 }& {\bf -53.
5 }
& {\bf 6.5 }& {\bf 0.57 }& {\bf 2.34 }      \\
(10385) & 23677 & BATSE 5648 & 40. & 59.4 & -52.6 & 3.3 & 0.84 & \\
& {\bf 24781 }& {\bf UC 10385c }& {\bf 49.} &{\bf 56.4 }& {\bf -53.0 }& {\bf 1.8
 }
& {\bf 0.67 }& {\bf 1.68 }\\
& 24350 & BATSE 5649 & None & 59.8 & -48.9 & 0.3  & None &\\
\multicolumn{4}{|r}{\bf 1300}&\multicolumn{5}{l|}{\bf P=0.007}\\
{\bf 961009 }& {\bf 49065 }& {\bf UC 10365c }& {\bf 20.} &{\bf 135.5 }& {\bf -79
.0 }
& {\bf 1.1 }& {\bf 6.69 }& {\bf 2.91 }     \\
(10365) & 49065 & BATSE 5629 & None & 130.2 & -80.2 & 0.4 & 6.4 & \\
&{\bf 50292 }& {\bf UC 10365d }& {\bf 36.} & {\bf 110.2 }& {\bf -79.0 }& {\bf 6.
6 }
& {\bf 0.25 }& {\bf 2.57 }\\
& - & - & - & - & - & - & - &\\
\multicolumn{4}{|r}{\bf 1150}&\multicolumn{5}{l|}{\bf P=0.2}\\
{\bf 961027 }& {\bf 42247 }& {\bf UC 10383d }& {\bf 75.} &{\bf 72.0 }& {\bf -43.
5 } & {\bf 10.4 }& {\bf 0.42 }& {\bf 3.04 }      \\
(10383) & 42247 & BATSE 5646 & 109. & 67.4 & -42.4 & 5.6 & 109. & 0.47  \\
& {\bf 43323 }& {\bf UC 10383e }& {\bf 2.} &{\bf 84.4 }& {\bf -51.0 }& {\bf 17.4
 } & {\bf 0.78 }& {\bf 8.2 }\\
& 43322 & BATSE 5647 & 1. & 68.7 & -54.3 & 5.8 & 0.85 & \\
\multicolumn{4}{|r}{\bf 1000}&\multicolumn{5}{l|}{\bf P=0.7}\\
{\bf 980213 }& {\bf 63928 }& {\bf UC 10857c }& {\bf 198.} &{\bf 11.1 }& {\bf -23
.8 }
& {\bf 7.4 }& {\bf 0.33 }& {\bf 2.23 }      \\
 (10857) & {\bf 64849 }& {\bf UC 10857d }& {\bf 104.} &{\bf 6.2 }& {\bf -10.6 }
& {\bf 13.1 }& {\bf 0.25 }& {\bf 2.7 }\\
\multicolumn{4}{|r}{\bf 950}&\multicolumn{5}{l|}{\bf P=0.6}\\
{\bf 960826 }& {\bf 57175 }& {\bf UC 10321d }& {\bf 45.} &{\bf 191.1 }& {\bf 15.
8 }
& {\bf 10.2 }& {\bf 0.32 }& {\bf 1.68 }      \\
(10321)& {\bf 58072 }& {\bf UC 10321e }& {\bf 34.} &{\bf 191.1 }& {\bf 28.3 }& {
\bf 12.6 }
& {\bf 0.23 }& {\bf 4.63 }\\
\hline
\end{tabular}
\end{center}
\end{table}

\begin{figure}
\includegraphics[width=8cm, height=8cm]{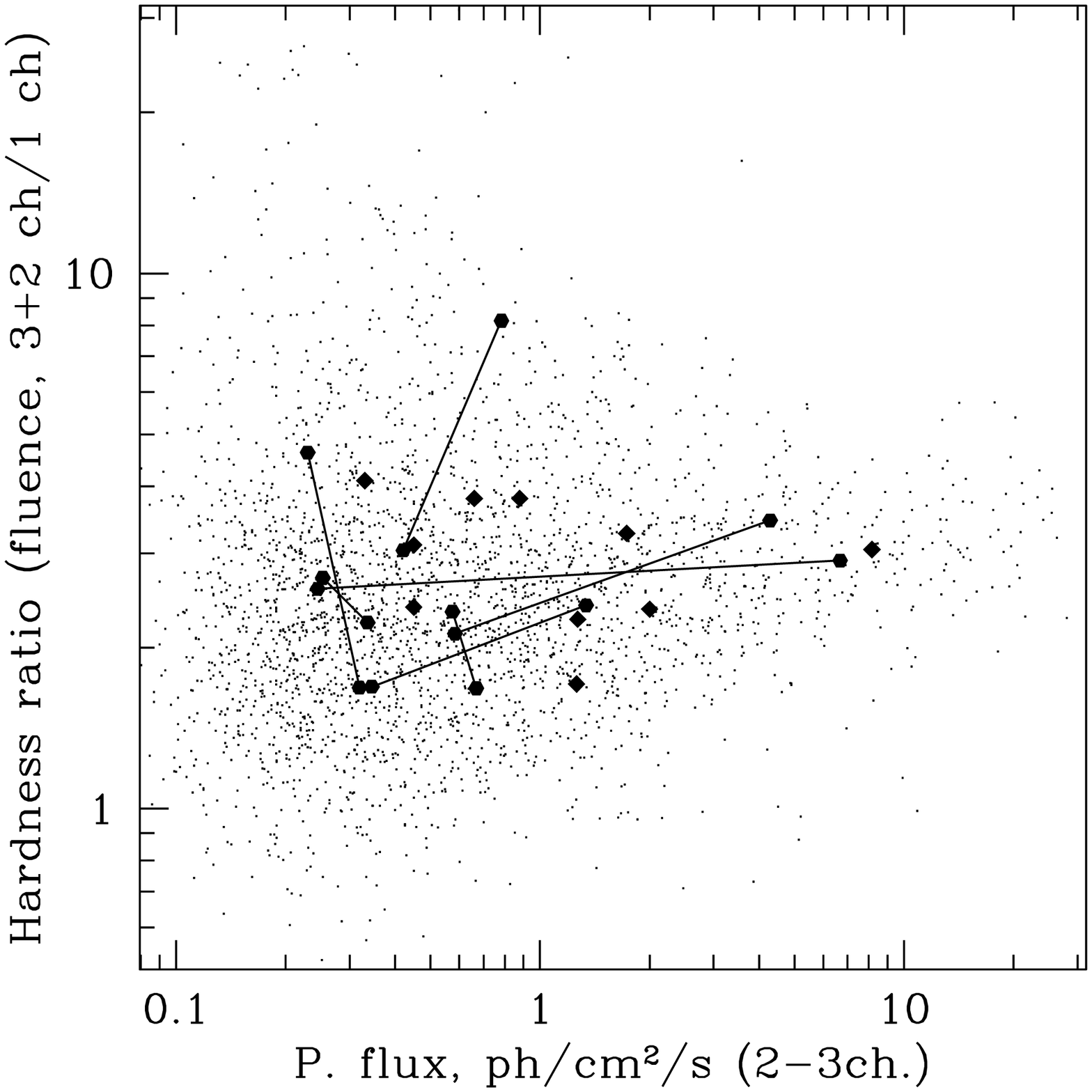}
\caption{Hardness of the GRBs found in the BATSE DISCLA records 
as the ratio of the count rates in the energy ranges 50--300 keV 
(spectral channels 2 and 3) and 25--50 keV (channel 1). 
The peak photon flux in the energy range 50--300 keV is 
along the horizontal axis; the diamonds indicate 
the obvious superlong events from Table 2; 
the dumbbells indicate the correlated pairs of events 
that are candidates for superlong bursts; 
the dots indicate GRBs from the catalog by 
Stern and Tikhomirova (2002).
}
\end{figure}

   It should be noted that the BATSE observational algorithm
was not intended for the detection of superlong GRBs:
from the detection time immediately after the accumulation
of limited information, data were
transmitted from the satellite to the Earth; 
during the data transmission, another event was recorded 
only if it was more intense than the previous one. 
Hence, the remote faint episodes of anomalously long bursts 
must have been lost, and it is possibly these events that 
we found in the BATSE records as pairs 910814 and 961009 (Fig. 2).

   Note that we searched for pairs of bursts as possible episodes 
of superlong bursts separated by a time interval of $\sim$2000 s. 
For a longer interval, the expectation of a coincidence of 
two independent bursts in the sample is $>$ 1 for most pairs, 
and no possible candidates for superlong bursts can be identified. 
However, since no obvious superlong GRBs 
(with almost continuous emission) with a duration $>$2000 s 
were found over the long period of BATSE observations, 
such superlong bursts do not exist, because a wide morphology 
of light curves is typical of any duration.

   Like obvious superlong bursts, no candidate pairs are 
identified in the class of long bursts by their parameters, 
except their duration (see Table 3). In general, 
their light curves (Fig. 2) reflect a wide morphology 
of the profiles typical of GRBs. These do not differ 
in hardness from typical GRBs either (Fig. 3).
The second episode in a pair can be either harder or softer, 
more or less intense. There is no reason to believe 
that the second episode in a pair can be produced 
by a different physical mechanism, as is 
the case for the afterglow.

   No emission was observed from these events 
in other spectral ranges. Given the broad distribution 
of the intrinsic GRB luminosities (Stern et al. 2002), 
it is difficult to draw a conclusion about the distance 
to the sources of these bursts from their brightness; 
however, these are clearly neither the nearest nor 
the farthest GRBs. If the time extension due to 
the unusually high redshift were responsible for 
the anomalous duration, this would affect the hardness 
of these GRBs, which is not observed.

   We present the duration distribution for all 
the long GRBs found in the BATSE DISCLA records, 
including the detected obvious superlong bursts and 
superlong candidate pairs (Fig. 4). Since breaks are 
regularly encountered in the records and a considerable part 
of the sky is shielded by the Earth when observing 
from a satellite, the probability that superlong bursts 
could be identified as such in the records 
(without the loss of episodes) is lower than that 
for moderate-duration bursts. Therefore, we estimated
the probability of identifying superlong GRBs
in the records by the Monte Carlo method. For
superlong bursts, we estimated the probability that
no less than 90\% of the dyration of their emission
will be seen in the records. The distribution of GRB
durations corrected for this probability declines
toward the longer durations, which roughly 
corresponds to a power-law  $dN/dT \propto T^{-1.5}$.

\begin{figure}
\includegraphics[width=10cm, height=8cm]{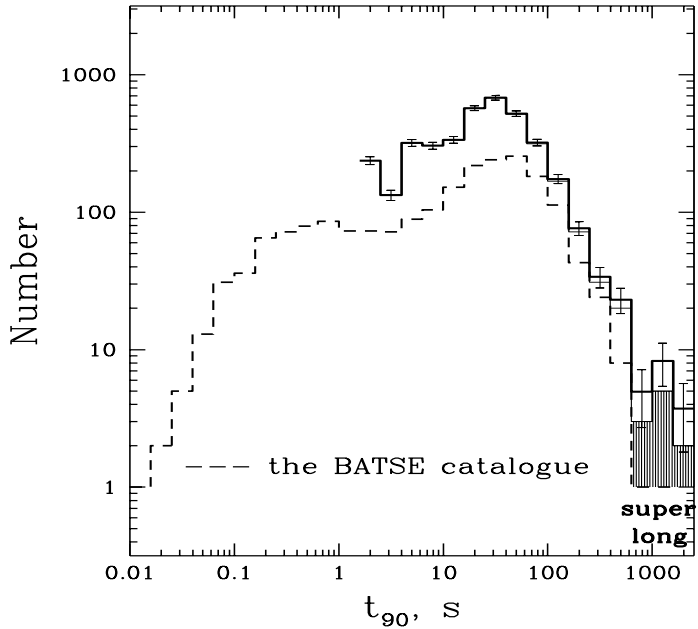}
\caption{Distribution of GRB durations. 
The thin line indicates the GRBs found in
the BATSE DISCLA records with the superlong bursts 
found here, including the correlated pairs; 
we use a visual estimate of the total duration 
for the superlong bursts and t90 for the remaining events; 
the thick line indicates the same distribution corrected 
for the probability of observing the total event duration; 
the dashed line indicates the distribution of durations 
according to the BATSE catalog.
}
\end{figure}

\section{CONCLUSIONS}

We scanned the BATSE continuous all-sky monitoring records
in the energy range 25 keV-1MeV for the entire observing
period of 9.1 yr and found ten superlong ($>$500 s) GRBs.
The two longest bursts have a duration of $\sim$1000-1360 s.
For four of the ten bursts, their duration is known from
the BATSE catalog; for the remaining bursts, we give
an estimate of their duration for the first time,
including that for one newly detected, nontriggered GRBs.
The fraction of obvious superlong ($>$500 s) GRBs 
among all of the long ($>$2 s) bursts found 
in the records is 0.3\%, which is slightly higher 
than the estimate that follows from the BATSE catalog. 
Superlong bursts do not differ in their main properties 
from long bursts.

   In addition, we detected bursts that could in pairs 
be episodes of superlong bursts with a duration of 
1000--2300 s. With these events, the fraction 
of superlong bursts in the class of long GRBs reaches 0.5\%. 
The existence of such long bursts ($\sim$2000 s) requires 
confirmation (or disproof) by future experiments 
with a high GRB localization accuracy. 
The extension of the GRB emission on time scales $>$2500 s 
is unlikely.

   The extension of the gamma-ray emission to time scales 
of $\sim$1000 s and father from the beginning of a GRB 
must be explained in terms of existing models and 
be taken into account when searching for 
an early X-ray afterglow.
   
\section*{ACKNOWLEDGMENTS}

   This work was supported by a 2004 grant 
from the Foundation of Support for Russian Science 
for young scientists, the Russian Foundation for 
Basic Research (project no. 04-02-16987), and a 
NORDITA grant (Nordic project in high-energy astrophysics 
in the INTEGRAL era.

\section*{REFERENCES}

\begin{flushleft}
1. V. Connaughton, Astrophys. J. 567, 1028 (2002).

2. V. Connaughton, M. Kippen, R. Priece, and K. Hurley,
IAUCirc6785, 1(1997).

3. G. J. Fishman, The Gamma Ray Observatory Science 
Workshop Ed. by W. N. Johnson (GSFC,
Greenbelt, 1989), p. 3.

4. K. Hurley, (2004) (private communication).

5. C. Kouveliotou, C. Meegan, G. Fishman, et al., 
Astrophys. J. 413, L101 (1993).

6. E. Mazets, S. Golenetskii, V. Il'Inskii, et al., 
Astrophys. Space Sci. 80,85(1981).

7. W. Paciesas, C. Meegan, G. Pendleton, et al., 
Astrophys. J. Suppl. Ser. 122, 465 (1999).

8. B. E. Stern, Ya. Tikhomirova, D. Kompaneets, et al.
Astrophys. J. 563, 80(2001).

9. B. E. Stern, Ya. Tikhomirova, M. Stepanov, et al.
Astrophys. J. 540, L21 (2000).

10. B. E. Stern, Ya. Tikhomirova, and R. Svensson, 
Astrophys. J. 573, 75 (2002).

11. J. van Paradijs, C. Kouveliotou, and A.M.J. Wijers,
Ann. Rev. Astron. Astrophys. 38, 379 (2000).
\end{flushleft}
\end{document}